# THE FACTORS OF THE CLASSIFICATION OF
# PROTEIN AMINO ACIDS

Miloje M. Rakočević

*Department of Chemistry, Faculty of Science, University of Niš,
Ćirila i Metodija 2, Serbia* (E-mail: m.m.r@eunet.yu )

**Abstract**.

In this work it is shown that three pairs of the factors appear to be the key, i.e. main factors of a natural classification of protein amino acids, i.e. canonical amino acids within the amino acid (genetic) code. First pair: the factors of the **habit** (habitus) of an amino acid molecule (size and polarity). Second pair: the factors of the **association** (type of the amino acid – enzyme reactivity and degree of the hydrophobicity–hydrophilicity of an amino acid molecule). Third pair: the factors of the **dissociation** (degree of the acidity-basicity, over acidic COOH group and degree of the basicity–acidity over the basic $NH_2$ group). As a result of the influence and interdependence of all six factors (measured through correspondent valid parameters) it appears still one natural classification into polar and non-polar amino acids, where polar amino acids possess negative and non-polar, the positive values of hydropathy index.

## 1. Introduction

Doolittle (1985, p. 76) has shown that a possible classification of protein amino acids (AAs) must be "based on the size of the amino acid's side chain and on the degree to which it is polarized". In this work, however, we start with a *working hypothesis* that three pairs of the factors appear to be the key, i.e. main factors of *a natural classification* of protein amino acids, i.e. canonical amino acids within the amino acid (genetic) code (Tables 1 & 2). First (the Doolittle's) pair: the factors of the **habit** (*habitus*) of an amino acid molecule [1. Size, in relation to the position, i.e. *topos* within genetic code ($t$ in Table 2) and 2. Polarity, i.e. hydropathy ($p$)]. Second pair: the factors of the **association** [1. Type of the amino acid – enzyme (aminoacyl-tRNA synthetases) reactivity ($e$) and 2. Degree of the hydrophobicity–hydrophilicity of an amino acid molecule ($h$)]. Finally, as a third pair: the factors of the dissociation [1. Degree of the acidity-basicity, measured over acidic COOH group within the amino acid molecule ($a$) and 2. Degree of the basicity–acidity, measured over the basic $NH_2$ group ($b$ in Table 2)][1].

In a proof proceeding of the working hypothesis, we start from the classification of protein amino acids into ten pairs (Rakočević & Jokić 1996; Rakočević, 1998)[2], as it is presented in Table 1 and Table 2. By this to the class of valine (in six different variants) belong AAs designated with "+" whereas to the class of glycine belong AAs designated with "−" (Table 2)[3].

---

[1] The classification into two classes (I and II in Table 1), accordingly with two classes of the enzymes aminoacyl-tRNA synthetases.

[2] Dlyasin (1998) also speaks about ten pairs. Eight pairs are identical by us and by him. The difference in two pairs (By Dlyasin: G-A/V-L, and by us: A-L/G-V) appears because Dlyasin does not include the classification of AAs into four stereochemical types (Popov, 1989; Rakočević & Jokić 1996).

| pK$_{COOH}$ | 2.3 | 1.8 | 2.3 | 2.4 | 2.4 | 2.2 | 2.2 | 1.8 | 2.4 | 2.2 |
|---|---|---|---|---|---|---|---|---|---|---|
| pK$_{NH2}$ | 9.6 | 10.0 | 9.2 | 9.7 | 9.6 | 9.7 | 9.1 | 9.0 | 9.4 | 9.1 |
| N$_2$ | 19 | 14 | 20 | 22 | 22 | 19 | 20 | 26 | 27 | 24 |
| N$_1$ | 10 | 05 | 11 | 13 | 13 | 10 | 11 | 17 | 18 | 15 |
| Vol. | 36 | 30 | 52 | 46 | 46 | 41 | 47 | 70 | 83 | 69 |
| Mass | 117.15 | 121.16 | 149.21 | 131.18 | 131.18 | 147.13 | 146.15 | 174.20 | 204.23 | 181.19 |
| H.ph. | 0.825 | 0.680 | 0.738 | 0.943 | 0.943 | 0.043 | 0.251 | 0.000 | 0.878 | 0.880 |
|  | + | + | + | + | + | − | − | − | − | − |
| I | V | C | M | I | L | E | Q | R | W | Y |
|  |  |  |  |  |  |  |  | | | | |
| II | G | S | T | P | A | D | N | K | H | F |
|  | − | − | − | − | + | − | − | − | − | + |
| H.ph. | 0.501 | 0.359 | 0.450 | 0.711 | 0.616 | 0.028 | 0.236 | 0.283 | 0.165 | 1.000 |
| Mass | 075.07 | 105.09 | 119.12 | 115.13 | 089.09 | 133.10 | 132.12 | 146.19 | 155.16 | 165.19 |
| Vol. | 03 | 21 | 32 | 31 | 14 | 30 | 36 | 58 | 50 | 62 |
| N$_1$ | 01 | 05 | 08 | 08 | 04 | 07 | 08 | 15 | 11 | 14 |
| N$_2$ | 10 | 14 | 17 | 17 | 13 | 16 | 17 | 24 | 20 | 23 |
| pK$_{NH2}$ | 9.8 | 9.2 | 10.4 | 10.6 | 9.9 | 10.0 | 8.8 | 9.2 | 9.2 | 9.2 |
| pK$_{COOH}$ | 2.4 | 2.1 | 2.6 | 2.0 | 2.3 | 2.0 | 2.0 | 2.2 | 1.8 | 1.8 |

**Table 1.** Two classes of protein amino acids. This Table shows ten pairs of amino acids, classified in accordance to Golden mean and to module 3:2 (Rakočević, 1998). Ten pairs within two classes of protein amino acids, the constituents of the genetic code [the sign "−" for polar and "+" for non-polar amino acids (after: Keyte & Doolittle, 1982 and Doolittle, 1986)]. Class I contains the larger amino acids (larger within the pairs), all handled by class I of enzymes, the aminoacyl-tRNA synthetases. Class II contains the smaller amino acids, all handled by class II of synthetases. Within the rows are given the values for the factors/parameters: hydrophobicity (H.ph.) and molecule mass (Both parameters after: Black & Mould, 1991); then Volume (Swanson, 1984), atom number $N_1$ as the number of atoms within the amino acid side chain, and $N_2$ as the number of atoms within the whole amino acid molecule; finely, the values for two constants: pK$_{COOH}$ pK$_{NH2}$ (Cantor & Schimmel, 1980). The correction of the pK$_{NH2}$ value for phenylalanine from 9.1 to 9.2 after: Pine, Hendrickson, Cram & Hammond (1980). Notice an interesting balance in the polarity: in relation to two identical AAs (both non-polar) A-L there are four non-identical pairs on the left. On the other hand, in relation to two non-identical AAs (one polar and other non-polar) F-Y there are four identical pairs (identical from the aspect of polarity) on the left.].

With the working hypothesis it is assumed that from the experimentally determined valid parameters (Table 1)[4], as a measure of the influence of the factors, follow the natural classes of amino acid molecules, such classes as aliphatic and aromatic compounds, chalcogenide derivatives (oxygen and sulphur derivatives), carboxylic amino acids and their amide derivatives etc. On the other hand, it must be that the polar and non-polar amino acid molecules also follow as separate classes.

---

[3] The valine class is identical with class I, and glycine class with class II in Tables 1 & 2 only in the case of the factor of amino acid - enzyme reactivity (column *e*).

[4] The belonging to two classes of amino acids in all cases is determined by degree – the major and less value of the parameter within the amino acid pair.

| I | t | p | h | e | b | a |   |   | a | b | e | h | p | t | II |
|---|---|---|---|---|---|---|---|---|---|---|---|---|---|---|---|
| **V** | + | + | + | + | + | + | 6:0 | 0:6 | − | − | − | − | − | − | **G** |
| **C** | + | + | + | + | − | + | 5:1 | 1:5 | − | + | − | − | − | − | **S** |
| **M** | + | + | + | + | + | + | 6:0 | 0:6 | − | − | − | − | − | − | **T** |
| I | + | + | + | + | + | − | 5:1 | 1:5 | + | − | − | − | − | − | **P** |
| L | − | + | + | + | + | − | 4:2 | 3:3 | + | − | − | − | + | + | **A** |
| **E** | − | − | + | + | + | − | 3:3 | 1:5 | + | − | − | − | − | − | **D** |
| **Q** | − | − | + | + | − | − | 2:4 | 2:4 | + | + | − | − | − | − | **N** |
| **R** | ± | − | − | + | + | + | 4:3 | 1:5 | − | − | − | + | − | − | **K** |
| W | + | − | + | + | − | − | 3:3 | 2:4 | + | + | − | − | − | − | **H** |
| Y | + | − | − | + | + | − | 3:3 | 3:3 | + | − | − | + | + | − | **F** |

**Table 2.** Three pairs of amino acid classification factors. The plusses and minuses designate the degree of the values ("+" more; "−" less) for the parameters: a. Acidity-basicity through $pK_{COOH}$ ("+" more acidic; "−" less acidic); b. Basicity- acidity through $pK_{NH2}$ ("+" more acidic; "−" less acidic); e. Amino acid relation to class I ("+") and class II ("−") of enzymes aminoacyl-tRNA synthetases; h. hydrophobicity (Black & Mould, 1991); p. Polarity for polar ("−") and nonpolar ("+") amino acids; t. "Topos" – the position within Watson-Crick Table: within inner three diagonals ("−"), and within outer four diagonals ("+"). Notice that within glycine class there is 66 molecules (66 minuses) and within valine class 55 molecules (55 plusses). If so, then the sum equals $121 = 11^2$, whereas the difference equals $11 = 11^1$.

## 2. TWO CLASSES OF AMINO ACIDS / ENZYMES

### 2.1. AA - enzyme reactivity and AA molecule size

All AAs within the row I in Table 1, and within column I in Table 2, are handled by class I enzymes aminoacyl-tRNA synthetases (plusses in column *e*), whereas all AAs within row/column II are handled by class II synthetases (minuses in column *e*). "The class I enzymes chare with dehydrogenases and kinases the classic nucleotide binding fold called the Rossmann fold, whereas the class II enzymes possess a different fold, not found elsewhere, built around a six-stranded antiparallel beta-sheet. The two classes differ … as to where on the terminal adenosine of the tRNA the amino acid is placed: class I enzymes act on the 2' hydroxyl whereas the class II enzymes prefer the 3' hydroxyl group." (Eriani et al., 1995 p. 499). On the other hand, the "class II synthetases are considered to be the more primitive of the synthetases" (Hartman, 1995, p 541).

Except to two classes of enzymes aminoacyl-tRNA synthetases, the plusses/minuses in column *e* are related to the molecule size from three aspects: molecule mass, volume and the number of particles within AA molecule (the number of atoms including the size of each atom, and the number of nucleons[5]; *see* legend to Table 1). In particular, that means that within each pair, the amino acid of class I is a larger, whereas of class II a smaller molecule. By this, bearing in mind that Class II of enyzmes is "more primitive" (Hartman, 1995, p 541), we can conclude that smaller molecules respond to more primitive enzymes, going from one to another AA pair (10 second members within 10 pairs).

---

[5] Only in the case of the pair C-S it exists, in the first slight, an uncertainty because both side chains possess the same atom number (5 atoms). However, the sulphur atom is larger than the oxygen atom.

## 2.2. A splitting within two AA classes by the module 3/2

The system of 20 canonical AAs, presented in Table 2 represents a perfect coherence between the physico-chemical properties of AA molecules and the balance of the number of atoms existing within the molecules. The balance appears to be self-evident if one can reveal the existence of a specifying splitting into two classes by module 3/2. [*Note about symmetry and harmony*. In a previous work (Rakočević, 1997), we have shown that the module 3/2 and/or 2/3 is the basic fractal motive within amino acid (genetic) code. The relation 2:3 and 3:2 are unique within the system of natural numbers series, because 2 is harmonic mean of the adherent number 3 and its half. On the other hand the number 3 represents a sum of the precursor 2 and its half. By all these relations one must bear in mind that the existence of a whole and its half is a condition for "the symmetry in the simplest case" (Marcus, 1989)]. The balance realizes itself through *four penthaplets*[6] and over a specific crossing (where the crossing exists also in the module 3/2) as follows. The number of atoms within 3+2 and 3+2 of AA molecules in upper half of the system differs for ±1 of atoms in relation to the arithmetic mean (first penthaplet VCM/PA with 39 -1 and second penthaplet GST/IL with 39+1 atoms). In lower half of the system within 3+2 and 3+2 molecules the difference is ±0 of atoms in relation to arithmetic mean (third penthaplet EQR/HF with 63±0 and fourth penthaplet DNK/WY with 63±0 of atoms also). Altogether, on the first zigzag line (invisible line, which connects AAs designated bold in Table 2) there are 102–1, whereas on the second zigzag line (AAs designated non-bold ) there are 102+1 of atoms. [*Note*. The number of nucleons on the first line is 628 + 10 and on the second line: 627 – 10. Cf. with nucleon number in Surveys 1-4. The nucleon number in all cases is as in Shcherbak (1993,1994), that means calculated only for the first nuclide within all five bioelements (H, C, N, O, S)][7]. In a previous paper, however, we have shown (Koruga, Rakočević et al, 1997) a different splitting into 10+10 of AA molecules, also in module 3/2, (and also with a crossing) with two zigzag lines, each line not with 102±1 but with 102±0 of atoms. By that, the vertically arranged amino acid doublets and triplets make not the horizontally arranged pairs.

## 3. The interdependence among position, size and polarity

The column *t (topos)* in Table 2 is related to position of AAs within Watson-Crick Table (WCT) of the genetic code. By the sign "–" are designated AAs which exist within inner space of WCT (**i**), that means within three middle diagonals, whereas by the sign "+" are designated AAs which exist within outer space of WCT (**o**), that means within two upper and two lower diagonals. The most AAs within outer space are non-polar (**n**) with the exceptions of Y, W and R, which are polar. On the other hand, the most AAs within inner space are polar (**p**) with the exceptions of F and L, which are non-polar.

---

[6] Notice that in the starting AA system (Table 2) we have four AA pentaplets, whereas at the ending AA system (Table 4; Section 4.2) – five quadruplets, what means a specific balance also.

[7] This arithmetical balance exists in a correspondence with chemical characteristics of AA molecules within two and two pentaplets. Therefore, within first two pentaplets, the doublets (PA and IL) are aliphatic or alicyclic AAs, whereas within second two pentaplets, the doublets (HF and WY) are aromatic AAs. The triplets existing within first two pentaplets (VCM and GST) contain two (ST) and two (CM) chalcogenide AAs, what means more diversity (OH and SH groups in relation to COOH or $NH_2$ groups). On the other hand, the triplets existing within second two pentaplets (EQR and DNK) contain two carboxylic AAs and their amide derivatives, plus one (R) and one (K) basic AAs. In first case in the 'game' is COOH group, and in second case $NH_2$ group, both existing within the AA 'head', what means less diversity. Finally, there are G-V in first two pentaplets without $CH_2$ group between the 'head' and side chain, and K-R in second two pentaplets with some $CH_2$ groups between the 'head' and side chain.

These strictly arranged positions of AAs within WCT (column *t* in Table 1: plus for outer and minus for inner AAs) appear to be in a strong correspondence with the polarity of AA molecule and its size (atom number) through the validity of the principle of minimum change as follows:

(n) 4V+1M+3I+4A+**2L**+**4L**+**2F**+2C = 22 molecules

    40+ 11+39+16+26 + 52 +28+10 = 2**2**2 atoms (42**0**)                                (1)

(o) 4V+1M+3I+4A+**2Y**+**4R**+**1W**+2C = 21 molecules

    40+ 11+39+16+ 30 + 68 +18+ 10 = 2**3**2 atoms (42**1**)                               (2)

(p) 4G+2K+2N+4P+**2Y**+**4R**+**1W**+2E+2D+4T+2R+2S+2Q+2H+4S = 39

    04+30+16+32+30 + 68+ 18+ 20 +14+ 32+ 34+10+22+22 + 20 = 3**7**2 (72**3**)   (3)

(i) 4G+2K+2N+4P+**2L**+**4L**+**2F**+2E+2D+4T+2R+2S+2Q+2H+4S = 40

    04+30+16+32+26+52 + 28+20+ 14+32+ 34 +10+22+ 22+ 20 = 3**6**2 (72**2**)   (4)

From the above presented equations follows that in 22 molecules of non-polar AAs there are 222 of atoms within their side chains, and 420 of atoms within the whole molecules; in outer space, however: 01 molecule less, 10 of atoms more within side chains and 01 atom more within the whole molecules. On the other hand, in 39 molecules of polar AAs there are 372 of atoms within their side chains and 723 of atoms within the whole molecules[8]; in inner space, however: 01 molecule more, 10 of atoms less within side chains and 01 atom less within the whole molecules. In reality, all these balances are directed by the "hidden" balances of the number of atoms within the AAs-exceptions in this manner: (2Y+1W = 48 atoms within side chains) + (4R = 68) = **7** molecules and 17**9** atoms; (2F = 28) + (6L = 78) = **8** molecules and 17**8** atoms.

## 4. THE NATURAL CLASSIFICATION OF AAS

### 4.1. The classification into polar and nonpolar AAs

The belonging of AAs to two classes (plusses for valine class and minuses for glycine class in Table 2) generates a new symmetric system correspondent with the natural numbers series: 3, 4, 5, 6 (Table 3)[9].

---

[8] Notice the cyclity of the permutations (372 and 723), which cyclity is found also by the balances in the nucleon number (by classification into four-codon and non-four-codon AAs) (Shcherbak, 1993, 1994).

| 6 | (−) | G | T | — | — |
|---|---|---|---|---|---|
| 5 | (−) | S | D | P | K |
| 4 | (−) | Q | N | H | — |
| 3 | (±) | E | Y | W | R |
|   |     |   |   |   |   |
| 3 | (±) | A | F | — | — |
| 4 | (+) | L | — | — | — |
| 5 | (+) | C | — | I | — |
| 6 | (+) | V | M | — | — |

| IV  | III | II  | I   |
|-----|-----|-----|-----|
| G-V | T-M | P-I | K-R |
| S-C | D-N | H-W |     |
| Q-E | Y-F |     |     |
| A-L |     |     |     |

**Table 3 (left).** Amino acid relations. The number of plusses and minuses follows from Table 2. The positions follow from the position in Table 2 as well as from the physico-chemical properties. Bold: polar, and non-bold: nonpolar amino acids.

**Table 4 (right).** Four Amino acids classes. If amino acids pairs must be as in Rakočević (1998) then four classes follow from the system presented in Table 3. Non-bold: amino acids of the alanine stereochemical type, i.e. the non-etalon amino acids; bold: etalon amino acids of glycine stereochemical type (G), proline stereochemical type (P) and valine stereochemical type (V, I). About etalon and non-etalon amino acids see in: Rakočević & Jokić (1996).

At the same time this generated system of plusses and minuses (as result of the influence of three pairs of the factors, i.e. parameters) exists in a strict interdependence with polar (bold in Table 3) and non-polar AAs. [*Note about plusses and minuses*: To be "positive," that means to be more "out," and to be "negative" that means to be more "in." Therefore, the larger molecules within class I in Table 1 are more "out" while smaller molecules within class II are more "in." The non-polar AAs (designated with plusses in Table 1) are more "out", that means near to the sphere from the aspect of electron density; the polar AAs are more "in", that means further from the said sphere. The more hydrophobic (less hydrophilic) AAs are more "out," that means not in contact with the water; the less hydrophobic (more hydrophilic) AAs are more "in," that means in close contact with water. The more acidic (less basic) AAs, from the aspect either COOH or $NH_2$ group, are more "out" because the hydrogen ion, i.e. proton is leaving; in contrary, the more basic (less acidic) AAs are more "in", that means designate with the sign minus].

### 4.2. The classification into five quadruplets and four classes of AAs

Bearing in mind that AA pairs are such pairs as we have been found within Class I and Class II in Tables 1 & 2, and knowing that a strict arrangement-order of AAs which we found in Table 3 is result of their physico-chemical properties, then unambiguously follows the arrangement-order of AAs given in Table 4.

Within first row of Table 3 must be only two AAs (with 6 minuses): G in first column and T in second column. Within second row there must be four amino acids (5 minuses): S in first column as a derivative of G (atom H is substituted with an OH functional group) and D in second column as an analog of T. [Aspartic acid as $HOOC-CH_2-CH(NH_2)-COOH$ and threonine as $HO-CHCH_3-CH(NH_2)-COOH$. Cf. side chains: in first case **HO-CO** and in second case **HO**; then H-C-H in first case and H-C-CH₃ in second case].

---

[9] One can notice here (for the sequence 3-4-5-6) the validity of the Pythagorean law ($3^2+4^2=5^2$) and Plato's law ($3^3+4^3+5^3 = 6^3$). About the fact that nucleon number within AA molecules (within four-codon AAs) is determined by the Pythagorean law, see in Shcherbak (1994, Fig. 1).

Further, it must follow P with **three** CH$_2$ groups and, finally, K with **four** CH$_2$ groups. Within third row there must be three AAs (with 4 minuses): N must go with D as its amide derivative; H with P (heterocyclic AAs) and Q with S. (The NH$_2$ amide group in glutamine comes as a result of a substitution of an OH group, which group exists in serine too).

If within first three row there are only polar amino acids, then within fourth row there must be also polar AAs – exactly four of them: E with Q (as carboxylic AA and its amide derivative), W with H (heterocyclic aromatic AAs) and R with K (two basic AAs). If so, then Y must go with N, D and T (the connection goes through OH function).

All other AAs are non-polar: A and F with 3 plusses, L with four, C and I with five, V and M with six plusses. If alanine, as first, must be in first column, phenylalanine must be in second column (together with tyrosine, which is its derivative). Leucine as aliphatic must be with alanine. Then comes C–I and V–M. Where? From pairing process, represented in Table 1, we understand that C must go into the column of S; I in column of P; V in column of G and, finally, M in column of T.

From above follows the next arrangement-ordering: one pair in fourth column in Table 3; one outer (P-I) and one inner pair (H-W) in third column; one outer (T-M) and two inner pairs (D-N and F-Y) in second column; two outer (G-V and S-C) and two inner pairs (E-Q and A-L) in first column. Altogether, an unambiguous arrangement-ordering as it is represented in Table 4.

Reading from a diagonal arrangement we find *four quadruplets* (see footnote 6) of 16 AAs of alanine stereochemical type in an original natural classification. Within first diagonal there are two quadruplets: aliphatic AAs (A-L and K-R), as 1$^{st}$ quadruplet and aromatic AAs (F-Y and H-W), as 2$^{nd}$. Within second diagonal are carboxylic AAs, i.e. carboxylic AAs and their amide derivatives (D-N and E-Q), as 3$^{rd}$ quadruplet. Finally, within third diagonal exist chalcogenide amino acids (S-C and T-M), as 4$^{th}$ quadruplet.

Bearing in mind that side chain represents a 'copy' of the AA 'head' we can speak about AAs which possess less diversity and about AAs which possess more of the diversity. Thus, the four quadruplets appear to be two octets: first octet with less diversity (aliphatic and carboxylic AAs, i. e. 1$^{st}$ and 3$^{rd}$ quadruplets, with 86-1 of atoms within their side chains); and second with more diversity (aromatic and chalcogenide AAs, i.e. 2$^{nd}$ and 4$^{th}$ quadruplets, with 86+1 of atoms within AA side chains)[10]. [*Note about diversity*. The 'copy' of a COOH or NH$_2$ (or NH) group means less diversity, whereas of an OH group means more diversity, because the OH group generates a very new type of organic compounds].

In a vertical-horizontal reading of the system presented in Table 4 we find four classes of AAs arranged hierarchically from two aspects. In first case, hierarchically from the aspect of the correspondence to the series of natural numbers 1, 2, 3, 4 in the next sense: 1 pair within first column, 2 pairs within second column, 3 pairs within third column and 4 pairs within fourth column. In second case, hierarchy follows from the aspect of the possession of a functional group. So, within 4$^{th}$ column we have less-more-less complex molecules. But within 3$^{rd}$ column it exists a vice versa situation: more-less-more complex molecules. On the other words, aliphatic pair A-L is less complex than aromatic F-Y; the carboxylic pair E-Q is more complex than D-N; finally, the chalcogenide S-C pair is less complex than T-M. [*Note*. In relation to this last point (T-M) there are two pairs of etalon amino acids, existing within the remaining three stereochemical types: G-V as less complex on the left, and P-I as more complex on the right].

---

[10] First octet with less diversity possesses less nucleons within AA side chains (506), whereas second octet with more diversity possesses more of nucleons (607). Notice that digital patterns of the nucleon number notations (506-607) is a noteworthy fact in the respect of discussed particles balances, the principle of minimum change and in an analogy with the quantum physics (607 – 506 = 0101)(Shcherbak, 1994, p. 476: "The laws of additive-position notation of numbers ... have analogies with quantum physics").

Within 2$^{nd}$ and 1$^{st}$ column, the functions realize themselves as more-more complex, the more complex aromatic H-W (more complex than aromatic F-Y) and the more complex aliphatic K-R (more complex than aliphatic A-L). With the comparison of 4$^{th}$ and 2$^{nd}$ column, one can understand the position of 5$^{th}$ quadruplet, the etalon quadruplet, with the functions less-less and more-more respectively. Namely, In the beginning of 4$^{th}$ column there is the less complex aliphatic pair (A-L), and at the end the less complex etalon pair (G-V). On the other hand, in the beginning of 2$^{nd}$ column, there is the more complex aromatic pair (H-W) and at the end, the more complex etalon pair (P-I). [About four stereochemical types and etalon AAs, see in Rakočević and Jokić (1996)].

**4.3. The splitting within five quadruplets by the module 3/2 (I)**

From the system, presented in Table 4, follow two step-by-step AA orders. The first, when the sequencing process begins with outer quadruplet from first diagonal (Model 1 in Survey 1); and the second, when the sequencing process begins with the inner quadruplet from the first diagonal (Model 2 in Survey 2). In Survey 1 it is shown that five quadruplets from Table 4 can be split into two groups through a crossing and by [2 x (3+2)] of pairs in each. In such a manner, all five quadruplets are splitting into two halves with 1:1 pair (reading vertically) and with a strict balance in the particles number: (102 & 102) atoms and (628 & 627) nucleons on each of two zigzag lines. In Survey 2, however, it is shown that five quadruplets can be splitting also into two groups by [2 x (3+2)] of vertical pairs, but into one inner and one outer group with (102±1) of atoms and (628 & 627) of nucleons within the groups.

**5. THE INTERDEPENDENCE BETWEEN POSITIONS AND FACTORS**

**5.1. A symmetric arrangement-ordering of AA quadruplets**

All classification and balances presented in four previous Sections represent interdependence between AA positions within WCT and the influence of three pairs of the factors. By this the word is about the positions determined by one and only distinction: to be within inner or within outer space in WCT (cf. the equations 1 – 4). In some way, this is very surprisingly knowing that it is possible to establish the positions for all canonical AAs within WCT, minimum in two manner: as in First Damjanović's system (DS1) – the positions from 0 to 19 (Fig. 2 in Damjanović, 1998) and as in Second Damjanović's system (DS2) (*personal comunication*) – the positions from 1 to 24 (Fig. 1 in this paper).

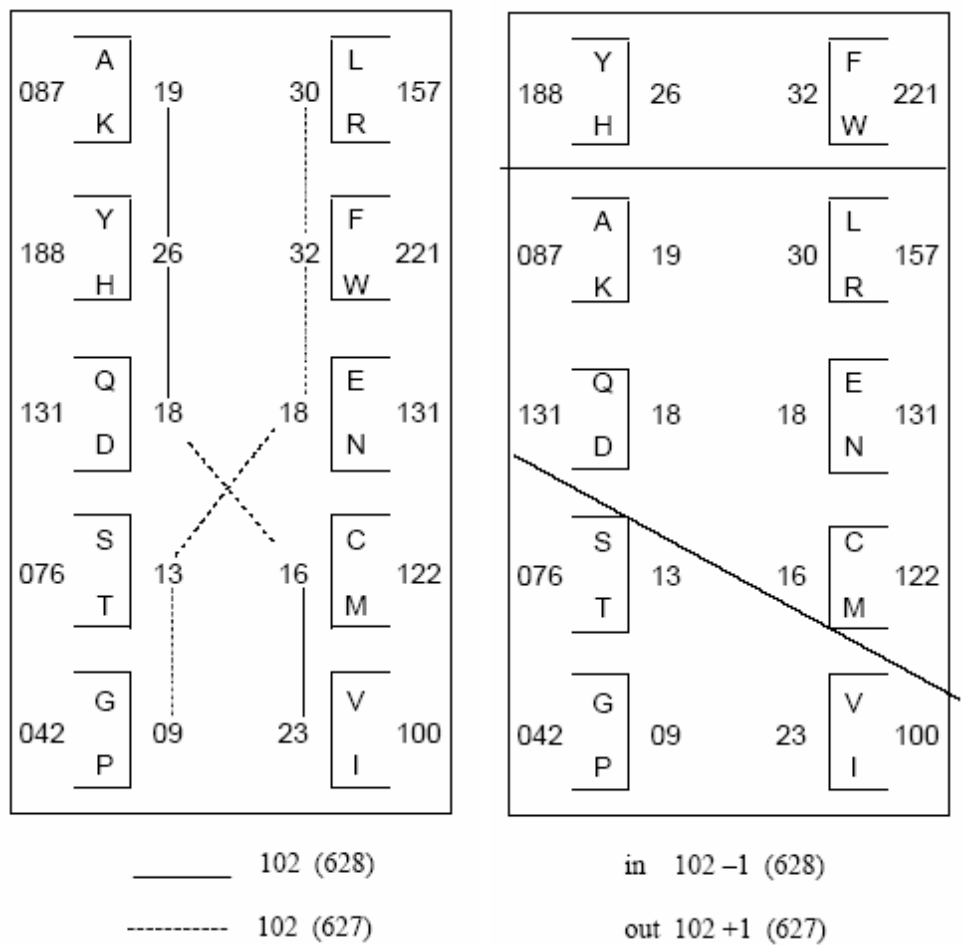

**Survey 1 (left).** The splitting of AA quadruplets (Model 1); **Survey 2 (right).** The splitting of AA quadruplets (Model 2)

In such a situation, it makes sense to research a possible interdependence among two Damjanović's systems and our system presented in Table 4. So, Table 5 represents the connection between DS1 and our system. [At Fig. 2/a in Damjanović (1998) the position number for an one-meaning AA, such as alanine (A) equals $12_4$ or $6_{10}$; for methionine (M) $30_4$ or $12_{10}$ etc. for other one-meaning AAs; the position number for a two-meaning AA, such as arginine (R), represents a middle value for two numbers: 1/2 of $(20_4 + 21_4)$ or 1/2 of $(08_{10} + 09_{10})$ etc. for other two-meaning AAs]. On the other hand, Table 6 represents the connection between DS2 and our system.

As we can see, the results of two connections are two very symmetrical systems, first in Table 5 and second in Table 6. The middle position in both possesses the quadruplet of aromatic AAs. Out of its there are 4±0 outer pairs within first and 4±1 within second system, all (outer) pairs arranged in four quadruplets. So, at the arrangement-ordering within first system (Table 5) the quadruplet of aliphatic AAs is above and quadruplet of carboxylic AAs (and their amide derivatives) below; then follow two quadruplets in middle positions: etalon AAs exactly very close to aromatic AAs, and one step further – the chalcogenide AAs.

| K | Q | E | * | N | H | D | Y |
|---|---|---|---|---|---|---|---|
| 1 | 2 | 3 | 4 | 17 | 18 | 19 | 20 |
| T | P | A | S |   |   |   |   |
| 5 | 6 | 7 | 8 |   |   |   |   |
| R | G | L | V |   |   |   |   |
| 9 | 10 | 11 | 12 |   |   |   |   |
| R | W | M | L | S | C | I | F |
| * |   | I |   |   |   |   |   |
| 13 | 14 | 15 | 16 | 21 | 22 | 23 | 24 |

**Figure 1.** The amino acid order after the second Damjanović's model (personal communication)

Within second system (Table 6), the arrangement-ordering realizes itself with one crossing step within four outer quadruplets. By this, carboxylic quadruplet is also below, but above exists a vice versa situation in relation to the system in Table 5: I-P, M-T, L-A versus L-A, M-T, I-P etc.

**5.2. The splitting within five quadruplets by the module 3/2 (II)**

For further comparison of systems DS1 and DS2 with our system, we must make a rearrangement of first system (Table 5) and of second system (Table 6); such a rearrangement in which on the top must be aromatic AAs and other quadruplets must go pair-by-pair and step-by-step as it is shown in Surveys 3 & 4 (Models 1' and 2', respectively). In the comparison of Models 1 & 1' and then of Models 2 & 2' we see that the balance of atom number is changed, accordingly to the principle of the minimum change: from $102 \pm 0$ to $102+1$, whereas the balance of nucleon number is not changed (628 & 627); in second case the balance is the same, except of a change in the direction: the state *in/out* within Model 2 is changed in the state *up/down* within Model 2'.

**5.3. The distinction by the module 3/2**

If we designate the full zigzag line with roman number I and dotted zigzag line with II, then very interesting relations follow. In the comparison of the subsystem 1(I) in Model 1 with the subsystem 2(in) in Model 2, we see that three pairs are the same and two different. The pairs A-K, Q-D and C-M appear in both subsystems. By this within the subsystem 1(I) exist still Y-H / V-I with 49 atoms and 288 nucleons, whereas within the subsystem 2(in) exist E-N / L-R with 48 atoms and 288 nucleons. On the other hand, in comparison of 1(II) with 2(out) we see that also three pairs are the same and two different. So, the pairs F-W, S-T and G-P appear in both subsystems. By this within subsystem 1(II) exist still L-R / E-N with 48 atoms and 288 nucleons, whereas within the subsystem 2(out) exist V-I / YH with 49 atoms and 288 nucleons.

| | | | | | |
|---|---|---|---|---|---|
| ⌈ R - K | (8.5 - 0.0 = 8.5) | | ─ I - P | (19.0 - 6.0 = 13.0) | |
| ⌊ L - A | (14.0 - 6.0 = 8.0) | | ⌈ M - T | (15.0 - 5.0 = 10.0) | |
| ⌈ M - T | (12.0 - 4.0 = 8.0) | | ⌈ R - K | (11.0 - 1.0 = 10.0) | |
| ⌈ I - P | (12.0 - 5.0 = 7.0) | | ⌊ C - S | (22.0 - 14.5 = 7.5) | |
| ⌈ H - W | (17.0 - 11.0 = 6.0) | | ⌊ L - A | (13.5 - 7.0 = 6.5) | |
| Y - F | (19.0 - 15.0 = 4.0) | | F - Y | (24.0 - 20.0 = 4.0) | |
| V - G | (14.0 - 10.0 = 4.0) | | H - W | (18.0 - 14.0 = 4.0) | |
| C - S | (11.0 - 7.5 = 3.5) | | ⌈ D - N | (19.0 - 17.0 = 2.0) | |
| ⌈ D - N | (18.0 - 16.0 = 2.0) | | ⌊ V - G | (12.0 - 10.0 = 2.0) | |
| ⌊ E - Q | (2.0 - 1.0 = 1.0) | | ⌊ E - Q | (3.0 - 2.0 = 1.0) | |

**Table 5 (left).** Five amino acid quadruplets (System I). In the middle are four aromatic amino acids. Above: four aliphatic, A-L as source aliphatic amino acids and K-R as amine derivatives. Below: carboxylic amino acids with their amide derivatives (altogether: four). Within the first concentric cycle: four etalon amino acids. Within second concentric cycle: four chalcogenide amino acids. The order as well as the numbers follow from the First Damjanović's system (Damjanović, 1998). Notice that in relation to aromatic amino acids, four pairs are above and four below (4±0 in relation to the arithmetic mean).

**Table 6 (right).** Five amino acid quadruplets (System II). All quadruplets as in Table 5, instead the order and numbers that follow from Second Damjanović's system (Figure 1). Notice that in relation to aromatic amino acids, five pairs are above and three below (4±1 in relation to the arithmetic mean).

However, with a cross in comparison occures a vice versa situation: two pairs of AAs are the same and three different. So, from the comparison of two subsystem 1(I): 2(out) distinctions are as follows: within the subsystem 1(I) A-K / Q-D / C-M with 52+1 atoms and 340 nucleons; within subsystem 2(out) G-P /F-W /S-T with 52-1 atoms and 339 nucleons. From the comparison 1(II): 2(in) the same distinctions follow, but in a vice versa order: within subsystem 1(II) there are G-P / F-W / S-T and within subsystem 2(in) there are A-K/Q-D/C-M.

Identical results can be obtained with a comparison of subsystems contained within Models 1' i 2' with only and one difference: instead the pair E-N the pair E-D appears, and instead the pair Q-D it appears the pair Q-N (the change for 1 atom and 1 nucleon).

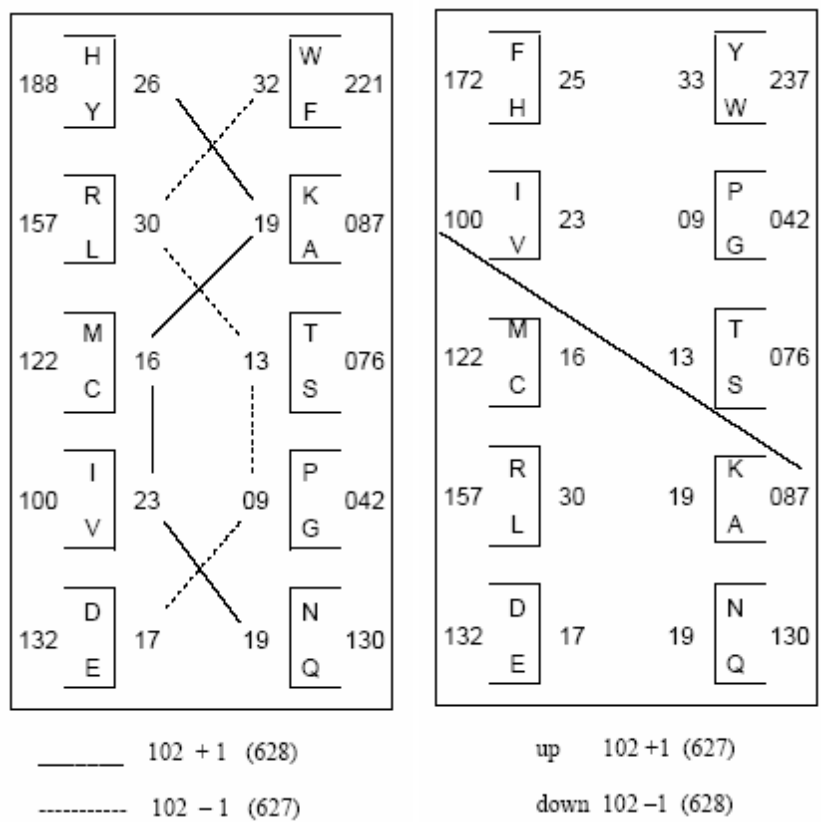

**Survey 3 (left).** The splitting of AA quadruplets (Model 1'); **Survey 4 (right).** The splitting of AA quadruplets (Model 2')

## 6. CONCLUSION

The argumentation given through five previous Sections of this paper provide evidence to support the working hypothesis, given in the Introduction, that three pairs of the factors, and corresponding parameters appear to be key determinants of a natural classification of protein amino acids. These three pairs unambiguously are: 1. Size and polarity of the AA molecule, 2. Type of the amino acid – enzyme reactivity and degree of the hydrophobicity–hydrophilicity of an amino acid molecule, and 3. Degree of the acidity-basicity, measured over acidic COOH group within the amino acid molecule, and degree of the basicity–acidity, measured over the basic $NH_2$ group. On the other hand, the natural amino acid classes are: 1. Aliphatic AAs, 2. Aromatic AAs, 3. Chalcogenide AAs and 4. Carboxylic amino acids and their amide derivatives – all four classes within alanine stereochemical type, and $5^{th}$ class etalon AAs (G-P, V-I)("etalon": after Rakočević and Jokić, 1996) existing within glycine stereochemical type, proline stereochemical type, and valine stereochemical type, respectively. Independent of this classification, from the influence and interdependence of all six factors, follows still one natural classification into two classes: 1. Polar AAs, and 2. Non-polar AAs (after positive and negative values of hydropathy index).

**ACKNOWLEDGEMENTS**

I am grateful to my professors Bojana Grujić-Injac and Zvonimir Damjanović, and to my colleges Radivoje Papović, Aleksandar Tančić and Djuro Koruga, also to my postgraduate and doctoral studies students Lidija Matija, Nataša Mišić, Tanja Andjelković and Mateja Opačić, for stimulating discussions and benevolent critique.